# Fault and Performance Management in Multi-Cloud Based NFV using Shallow and Deep Predictive Structures

Lav Gupta
Dept. of CSE
WUSTL
St. Louis, USA

M Samaka
Dept. of CSE
Qatar University
Doha, Qatar

Raj Jain
Dept. of CSE
WUSTL
St. Louis, USA

Aiman Erbad
Dept. of CSE
Qatar University
Doha, Qatar

Deval Bhamare
Dept. of CSE
Qatar University
Doha, Qatar

H Anthony Chan
Huawei Technologies
Plano, TX
USA

*Abstract*—Deployment of Network Function Virtualization (NFV) over multiple clouds accentuates its advantages like flexibility of virtualization, proximity to customers and lower total cost of operation. However, NFV over multiple clouds has not yet attained the level of performance to be a viable replacement for traditional networks. One of the reasons is the absence of a standard based Fault, Configuration, Accounting, Performance and Security (FCAPS) framework for the virtual network services. In NFV, faults and performance issues can have complex geneses within virtual resources as well as virtual networks and cannot be effectively handled by traditional rule-based systems. To tackle the above problem, we propose a fault detection and localization model based on a combination of shallow and deep learning structures. Relatively simpler detection has been effectively shown to be handled by shallow machine learning structures like Support Vector Machine (SVM). Deeper structure, i.e., the stacked autoencoder has been found to be useful for a more complex localization function where a large amount of information needs to be worked through to get to the root cause of the problem. We provide evaluation results using a dataset adapted from fault datasets available on Kaggle and another based on multivariate kernel density estimation and Markov sampling.

## I. Introduction

Virtualization of network services using NFV can transform a network service provider's business model in many ways. By consolidating appliances and middleboxes onto commercially available, high volume servers, they reduce the time to market new services, provide flexibility of scaling and lower the cost of operation [1]. Cloud technology can multiply the benefits of NFV [2], [3]. It could provide greater flexibility in obtaining resources, bring Network Service Provider's (NSP's) points of presence close to customers, provide an opportunity to optimize performance and control cost. Deploying virtual resources on clouds, especially multiple clouds, is becoming a key to successful deployment of large real-time and distributed services. However, in this NFV on cloud scenario, concerns about five nines availability (99.999%) and quality of service parameters like latency and packet loss still remain [5], [6].

Traditional networks have rigorous availability and quality control. They have time-tested standards relating to fault, configuration, accounting, performance and security (FCAPS) as embodied in ISO Common Management Information Protocol (CMIP) and ITU TMN M.3010 and M.3400 recommendations [7], [8]. As against this, even though the complexity of NFV implementations is many notches higher, they still require serious work to become carrier grade and this is not going to be trivial [11], [12].

In this paper, we propose mechanisms to handle manifest and latent fault and performance issues in NFV over a multi-cloud environment. Our work shows that a combination of shallow and deep machine learning architectures would be useful in handling voluminous operational data of high dimension for detection and localization of fault and performance issues. The rest of the paper is organized as follows: Section II discusses how network services are organized in the target environment. In Section III we discuss the extent and complexity of the FCP part of the FCAPS problem and state the aspects we propose to handle in this paper. In Section IV we bring out state of the art through related works. Section V deals with the solution strategies while Section VI presents the evaluation results. In Section VII we present the summary.

## II. Network Service Structure

Based on the ETSI specifications [9] and IETF RFC [13] a network service can be described as an ordered set of virtual network functions (VNFs) that represent functions like routers, broadband network gateways or middleboxes like load balancers and firewalls. These functions are chained into service function chains (SFC) or VNF graphs interconnected by virtual network resources to handle the traffic in a desired way. VNFs are created in software and hosted on virtual machines (VM) in one or more clouds [10]. Creating virtual network service on multiple clouds gives a number of compelling advantages like lower capital and operational cost,

This work has been supported under the grant ID NPRP 6 - 901 - 2 - 370 for the project entitled "Middleware Architecture for Cloud Based Services Using Software Defined Networking (SDN)," which is funded by the Qatar National Research Fund (QNRF) and by Huawei Technologies. The statements made herein are solely the responsibility of the authors.

flexibility of resource selection, scaling and descaling, closer points of presence and faster time to market. Fig 1 shows an end-to-end network SFC using resources from cloud service providers (different domains).

Fig. 1 Multi-domain End-to-End Service

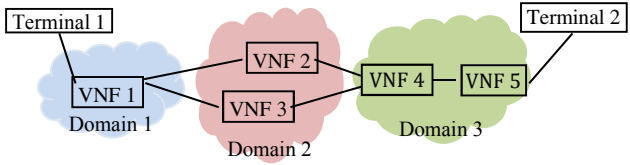

VNFs and SDCs are created and managed by the NFV Management and Orchestration (NFV-MANO) [14] on VMs that are managed by management and control platform (MCP) of the Cloud Service Provider (CSP). The ISP service (e.g., broadband) is managed by the Business Support System (BSS)/Operation Support System (OSS) owned by the ISP. Effective coordination of CSPs' MCP, ISPs BSS/OSS and the NFV-MANO is key to successful virtual network service delivery. Fig 2 shows the interfaces between the service provider and the NFV domains as defined by ETSI.

Fig. 2 Virtual Network Service and its Management

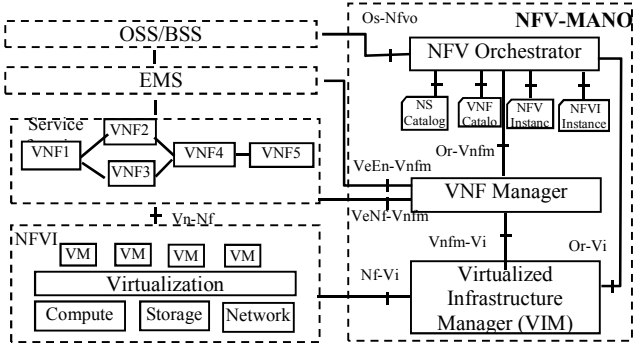

As far as the fault and performance management are concerned, the VNF Infrastructure Manager (VIM) interacts with the NFVI for collecting and forwarding performance measurement and events. The VNF Manager (VNFM) obtains device level statistics, logs notifications, alarms and events from the respective EMSs. NFV Orchestrator (NFVO) interacts with the BSS/OSS for configuration and fault management of network services.

### III. FCP PROBLEM DESCRIPTION

Faults may occur for a number of reasons, prominent of which are malfunctioning or failed devices because of hardware or software failures in VM or VNF, failure of links and configuration errors. There could be other reasons like cyber attacks, disasters or environmental factors. Faults appear as errors. Errors in turn are deviations of a system from normal operations. Errors are reported through system alarms. Alarms are notifications about specific events that may or may not be errors. Four levels of severity of events and alarms have been defined in ITU standard X.733: Critical, Major, Minor, and Warning [15]. The critical alarm appears when the service can no longer be provided to the user. Major alarm indicates the service affective condition while minor means no current degradation is there, but if not corrected may develop into a major fault. A warning is an impending service affecting fault or performance issue. To be able to make use of the multi-cloud paradigm effectively, it is extremely important that the fault and performance management issues are fixed for this environment [16]. The FCP system should be able to identify which issues are potential performance hazards or may result in a fault that would require resources to rectify.

There are a number of reasons why the combination of multi-cloud and NFV need a strong fault and performance management system. Cloud reliability today does not measure upto the network services' need of five nines. Detection and localization of faults and performance issues has a higher degree of complexity here as the anomalous behavior could be in the hardware, virtual machines, VNFs, SFCs or at the service levels. This involves coordination among management systems of the CSP, ISP and the NFV provider. The inter-se responsibilities have not been clearly demarcated. There are gaps in definition of multi-domain NFV-MANO coordination with the BSS/OSS and its effect on fault management [17]. ETSI document [14] speaks of including fault correlation and root cause analysis in the NFV system operation and maintenance. Fault escalation from lower to higher layers has not been defined. ETSI supported proof of concept (POC#35) has shown the shortcomings of the OpenStack MCP's ability in managing faults in a mission critical application.

Communication networks are widely distributed and are complex. The variety of FCAPS issues that can afflict them is large. To detect, diagnose and localize any condition that degrades network performance becomes quite onerous [20]. In this paper, we restrict ourselves to the fault and performance issues and make reference to configuration problems in as much as they are relevant to these. For the purpose of this paper, we could explain the FCP problem in terms of the following:

1. Detect and notify, manifest and impending, faults and performance issues that could be the cause of performance degradation or failure.
2. Identify and localize manifest and impending faults and performance problems. In case of impending faults severity level should be predicted.

### IV. STATE-OF-THE-ART

The ETSI documents relevant to FCAPS are the NFV resiliency requirements [21] and service quality metrics [22]. The former provides a list of faults and relationship between them while the latter gives VNF related metrics useful for quality of service. Some issues relating to FCAPS have not yet been dealt with. To begin with, the metric list needs to be supplemented with issues related to specific network services, e.g., 'continuous dial tone' or 'line card fault' or 'phone dead' in relation to fixed phones and similarly 'call drops', 'weak signal strength' and 'roaming failure' for mobile network and, 'training light flashing', 'WAN light flashing' or 'low data rate' for DSL broadband networks. Secondly, much is being left to the NSP's OSS/BSS and the "Os-Nfvo" interface (see

Fig. 1) which has been loosely defined. How the placement, migration, and reassignment are to be done has not been defined and left to implementers [23].

The standardization work has been fragmented among the industry standards bodies, including ETSI, 3GPP, Broadband Forum, IETF, ITU-T SG 15, MEF, ONF, OPNFV and TM Forum [24]. The OPNFV industry group is working on a project called 'Doctor' for fault management and maintenance that aims for high availability of network services on top of virtualized infrastructure. It is based on OpenStack telemetry and alarming and is not geared up for multiple clouds [25]. The current problem with OpenStack is its development being driven primarily by IT community resulting in lack of some critical capabilities for NFV. Use of shallow architectures in machine learning has been reported in a number of industrial settings like loss of performance in industrial plants using support vector machine (SVM) [26], [27], [28]. Deep learning architectures have been lately applied e.g., deep neural network based fault diagnosis in [29], and [30].

However, work on fault and performance management of virtual network services has been scarce. Some examples are: Artificial Neural Networks (ANN) for single and double alarm simulated scenarios [31], a system for fault analysis and prediction in telecommunications access network for Rijeka area of Croatia [32], fault prediction in telecommunications networks [33].

## V. FCP SOLUTION COMPONENTS

The virtualized systems require features that are either not present in the traditional diagnostic methods or not fully exploited. Prediction of impending failure, dealing with incomplete information and analyzing trends to predict failure are some of the key requirements. Modern communication systems produce large volumes of high-dimensional operational data. In such a case, analyzing the data to get an actionable understanding of the situation becomes difficult. It would be extremely difficult, if not impossible, to capture intricate relationships between the features and the labels through the traditional methods. In general, the researchers agree on predictive approaches that take a learning route to solve the problem of the complex interaction of features of fault detection and localization [34]. We have worked on shallow as well as deep structures in machine learning in an attempt to tackle this problem. We feel that a model based on a judicious combination of these could be used for prediction of faults and performance issues along with the severity levels of impending faults with a high level of accuracy.

### A. Model for Fault Detection and Localization

The proposed model, shown in Fig. 3, has predictive and deductive properties to meet the FCP requirements. The detection system first decides whether there is a manifest or an impending fault or a performance issue. Based on this, the system will launch into identification and localization. Failure prediction needs to be accompanied with a high probability of correctness as actions following such a prediction involve cost. For localization, the model uses a multi-layered strategy. First, the broad category of the fault is determined. The system then does finer grain localization within the broad category. For the impending faults, it also gives location and severity level of the fault.

Fig 3. Fault/Performance Detection and Localization Model

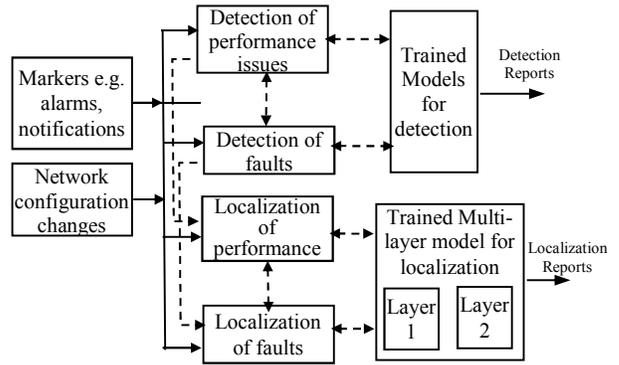

### B. Markers and Metrics for Fault Detection and Localization

There are events relating to communication, quality of service, processing, equipment and environment that produce alarms, notifications, warning or error messages, measurements, counter values and conditions. Of course, many of the markers will appear in more than one type of fault or performance issue. Once trained the detection and localization algorithms would be able to pick out relevant markers and use them to predict the type of condition that may have arisen. Some of the markers related to mobile, fixed and broadband networks are given in Table 1.

TABLE 1 LIST OF MARKERS

| Mobile Network | Fixed Network | Broadband |
|---|---|---|
| C/I ratio | No dial tone | Intermittent connection |
| Radio Link Time Out | Channel Noisy | Low Data Rate |
| Time Slot Shortage | MDF Jumper Dis | Phone Works Broadband Down |
| Occupied Bandwidth | Line Card Port Faulty | Repeated Training |
| RX Noise Floor | Primary Cable Fault | LAN Lamp Off |
| Radio Power | Distribution Cable Fault | Line Noisy |
| Frequency Error | DP Fault | DSLAM Port Mismatch |
| Antenna Tilt | House Wiring | No Ping |
| Signal Strength | MDF Fuse Blown | ADSL Lamp Flashes/Off |
| BTS Down | Customer Instrument Fault | No Line Sync |
| Handover Failure | Dis In One Limb | Browsing Issues |
| Roaming Failure | Earth Contact | Micro-Filter Faulty |
| Packet Loss | Drop Wire Fault | No Comms |
| Hypervisor Alarm | Ring Tone Fault | Dropouts |
| Registration Failure | Message Fault | No Authentication |
| Low CSSR | Delayed Dial Tone | |

Table 2 shows two common problems in cellular mobile networks in which markers overlap.

TABLE II FAULTS AND MARKERS

| | Phase Error | Power | EVM | Rx Noise floor | Origin offset | Occupied BW | Frequency Error | C/I |
|---|---|---|---|---|---|---|---|---|
| Call Drop[1] | Y | Y | Y | Y | Y | | Y | Y |
| Call Blocked[2] | | Y | Y | Y | Y | Y | | |

[1]Radio link timeout [2]Time slot short, EVM: Error Vector Magnitude, Rx: receiver, C/I: carrier to interference ratio, Y: marker present

ETSI documents on service quality [21] and [22] on reliability mention metrics that need to be collected and analyzed. The ETSI group specification on service quality metrics recognizes that it is important to have an objective and quantitative metrics for good service to the consumers, fault localization and identification of the root cause of performance deviations. Examples of metrics and their realistic values (where applicable) from an actual network are given in Table 3.

TABLE III METRIC FOR NETWORK AVAILABILITY AND RESILIENCY (NOT EXHAUSTIVE)

| Metric | Typical Value | Metric | Typical Value |
|---|---|---|---|
| **Mobile Network** | | | |
| **Network Availability** | | **Broadband Network** | |
| BTS total downtime | ≤ 2% | Packet loss | <1% |
| **Connection Establishment** | | ISP-POP to IGSP/IXCH Latency | <120ms |
| CSSR (call set up success rate) | ≥95% | ISP G/W to International NAP (terrestrial) | <350ms |
| SDCCH/Paging channel congestion | ≤ 1% | **Throughput/Bandwidth Utilization** | |
| TCH Congestion | ≤2% | POP to ISP G/W | <80% |
| **Connection Maintenance** | | Avg. throughput for packet data | >90% |
| CDR (call drop rate) | ≤ 2% | Latency (Audio) | <150ms |
| POI congestion | ≤ 0.5 | | |
| Signal Strength in vehicle | ≥85dBm | | |
| **Fixed Network** | | | |
| Fault incidences | <5% | | |
| CCR (call completion ratio) | >55% | | |
| IGSP: Internet Gateway Service Provider, IXCH: Internet Exchange, SDCCH: Standalone Dedicated Control Channel, TCH: Traffic Channel | | | |

*C. Training Datasets*

The quality and quantity of the datasets affect the learning and prediction performance of machine learning algorithms [69]. Information about faults, observations and restoration details in the telecommunication networks is contained in the fault dockets, test reports, central office system logs, outdoor maintenance staff logs, cable maintenance staff diaries and docket closure reports. Publicly available datasets like UCI repository [35], Stanford SNAP [36], DMOZ [37] have datasets relating to faults in contexts other than telecommunications networks. A number of authors [32] have either used proprietary datasets not publicly available or synthetic data sets.

We have used two datasets for evaluating our algorithms. One of the datasets pertains to actual faults and their severity levels in the Telstra network disruptions and the other is a synthetic dataset generated through multivariate kernel density estimation (KDE) technique augmented with real faults from a live network. The Telstra datasets (2016) are derived from the fault log files containing real customer faults [38] and adapted for this study. The training dataset (Table IVA) contains location and a time point. They are identified by the "ID" column, which is the key "ID" used in other data files. Fault severity has three categories with 0 indicating no fault, 1 indicating a few faults and 2 indicating many faults. There are datasets for event type, the features logged, the resource affected and the severity type. The severity_type is different from fault severity and classifies the warning given by the system. Table IVB gives an example of one of these datasets (resource type).

TABLE IVA TRAINING DATASET

| Id | Location | Fault severity |
|---|---|---|
| 14121 | Location 118 | 1 |
| 9320 | Location 91 | 0 |
| 14394 | Location 152 | 1 |
| 8218 | Location 931 | 1 |
| 14804 | Location 120 | 0 |
| 1080 | Location 664 | 0 |
| 14964 | Location 704 | 2 |
| 12896 | Location 613 | 2 |
| 7288 | Location 846 | 2 |
| 13300 | Location 613 | 1 |

TABLE IVB THE RESOURCE TYPE DATASET

| Id | Resource_type |
|---|---|
| 6597 | Resource_type 8 |
| 8011 | Resource_type 8 |
| 2597 | Resource_type 8 |
| 5022 | Resource_type 8 |
| 6852 | Resource_type 8 |
| 5611 | Resource_type 8 |
| 14838 | Resource_type 8 |
| 2588 | Resource_type 8 |
| 4848 | Resource_type 8 |
| 6914 | Resource_type 8 |

Generation of high dimensional synthetic data can be done using kernel density estimation with Markov sampling. This type of data is commonly employed as a substitute for real data and can be used to provide a controlled testing environment that meets specific conditions [39]. We have used the kernel density estimator for high dimension [39]. We call this the KDE dataset (Table V).

TABLE V FEATURES AND CLASSES IN THE KDE DATASET

| | Features | | Features | | Classes |
|---|---|---|---|---|---|
| 1 | BTS hardware | 14 | POI Congestion | 1 | Call drop |
| 2 | Radio link phase | 15 | Temperature | 2 | Call setup |
| 3 | EVM | 16 | Hypervisor | 3 | No Roaming |
| 4 | C/I ratio | 17 | HLR | 4 | Weak Signal |
| 5 | BSIC fault | 18 | VLR | 5 | No registration |
| 6 | BCC fault | 19 | Billing | 6 | No outgoing |
| 7 | Time slot short | 20 | MS | 7 | Data not working |
| 8 | Power | 21 | Virtual resource | | |
| 9 | Rx noise | 22 | ID Signal Strength | | |
| 10 | Antenna tilt | 23 | OD Signal Strength | | |
| 11 | Occupied BW | 24 | Handover | | |
| 12 | Filter fault | 25 | CSSR | | |
| 13 | TCH Congestion | 26 | SDCCH Congestion | | |
| EVM: Error Vector Magnitude, C/I: carrier to interference, BSIC: BCC, TCH: Traffic Channel, Rx: Receiver, SDCCH: Standalone Dedicated Control Channel, POI: Point of Interconnection, HLR: Home Location Register, VLR: Visitor Location Register, MS: Mobile Station, ID: Indoor, OD: Outdoor, CSSR: Call Setup Success Rate | | | | | |

*D. Discussion on Shallow and Deep Learning Methods Used*

In the context of this paper, we use the terms shallow structure and shallow architecture interchangeably and the same is the case with deep structure and deep architecture. Shallow structures are simpler with one stage of non-linear operation, e.g., one hidden layer in neural networks. Deep learning architectures would have more than one level of the composition of non-linear operations in the function learned. Deep neural networks have finally attracted widespread attention, mainly by outperforming alternative machine learning methods such as SVM in numerous important applications [40], [41]. One of the key advantages of deep learning is the automatic extraction of high-level features from the given dataset. This is a distinct advantage over the difficult feature engineering in shallow structures that that requires human intervention. In deep learning higher-level features are learned as a composite of lower level features. In this way, features are learned at many levels of abstraction making it easier to grasp complex functions that map the input to the output directly from data. As our model uses both shallow and deep structures we discus here briefly the important aspects of the algorithms that we have used in our study leaving the details to the references mentioned therein.

*1) Support Vector Machine (SVM):* The SVM for classification (SVC) is a supervised learning method that analyzes data and recognizes patterns. An SVM classifier produces an optimal hyperplane that separates different classes of data in the given labeled training data. The decision function is fully specified by a subset of training samples, the support vectors that lie closest to the hyperplane. This model can then be used to predict the class of new data points. Reference is made to [42] for a detailed description.

*2) Alternating Decision Trees (ADT):* This method is a variation of the standard decision tree that partitions instance space into disjoint sets by splitting the leaf nodes. Whereas in standard decision trees the leaf node is split once, in ADT it can be split multiple times. Boosting brings performance enhancements. More details are available at [43].

*3) Random Forest:* The Random Forest (RF) is a useful classifier that gives good results in many situations. It can be run efficiently with large data sets. It can also produce estimates of the relative importance of the predictor variables. It does not need a separate test dataset as the OOB-error gives an unbiased estimate of test or classification error [44].

*4) Deep Learning Through Stacked Autoencoders:* An autoencoder is a neural network consisting of two parts, an encoder function $h=f(x)$ and a decoder that produces reconstruction $r=g(h)$. The encoder maps input data to a lower dimension or compressed feature representation while the decoder converts it back to the dimensions of the input space. Because the model is forced to prioritize which aspects of the input should be copied to the output it often learns useful properties of the data. A stacked autoencoder consists of multiple layers of the sparse autoencoder in which the output of each layer is wired to the inputs of the successive layer. Learning an under complete representation (with $h$ having a smaller dimension than $x$) forces autoencoder to captures the most salient features of the training data. In the learning process the loss function, $L(x,g(f(x)))$ is minimized, where $L$ is the loss function and there is a penalty on $g(f(x))$ for being dissimilar from input $x$. In a sparse autoencoder each layer is pretrained to learn an under complete or sparse representation to reduce training time and improve accuracy. The training criterion then involves a sparsity penalty $\Omega(h)$ in addition to the reconstruction error and can be written as $L(x,g(f(x)))+\Omega(h)$. Where $\Omega(h)$ can be thought of as a regularizer term.

If we consider a stacked autoencoder with $n$ layers then $h^{(n)}$ is the activation of the deepest layer of the hidden units, which gives us a representation of the input in terms of higher-order features. For classification, the features from the stacked autoencoder can be used by feeding $h^{(n)}$ to a softmax classifier. Softmax is a regression method that can work with multiple classes.

The stacked autoencoder is trained in a greedy layer-wise manner. When all the layers have been trained fine-tuning is done using back propagation to tune the parameters of all the layers and produce better results [45].

### E. Detection of Fault and Performance Conditions

Faults and performance issues may range from simple single point failures to multiple correlated or uncorrelated events. A fault presents itself in the form of system malfunction and notifications from faulty and other connected devices. The failure detection mechanism should be able to filter out dependent and routine operational events so that resources are not wasted in localizing these problems. In NFV we deal with, VM, VNF and virtual network faults that cause virtual network services to behave abnormally. For example, failure of a Gigabit Ethernet interface on the core router may cause some or all of the virtual private network (VPN) links of many customers to be non-functional. In the context of this paper, the goal of the FCP detection mechanism is to correlate alarms, notifications, measurements and other markers generated by events to infer manifest or predict impending performance and fault conditions.

Some errors may be cleared by the system, others may produce warnings that may signal impeding problems while still another may produce faults that bring down functionalities and make themselves evident. In our implementation, the trained shallow machine-learning models learn from the past events relating to faults and their resolutions. The models work in two stages: the first stage just makes a decision between 'fault' and 'no-fault' conditions, while the second stage does a more detailed examination of the markers to choose between 'manifest' and 'impending' faults. Minor faults and warnings discussed above would be the main contributors to the impending faults and need to be analyzed to make this decision. With correct segregation the localization stage would be able to carry out its functions properly. Fig 4 shows the detection process (boxes for localization are not part of detection).

Fig 4 The detection process

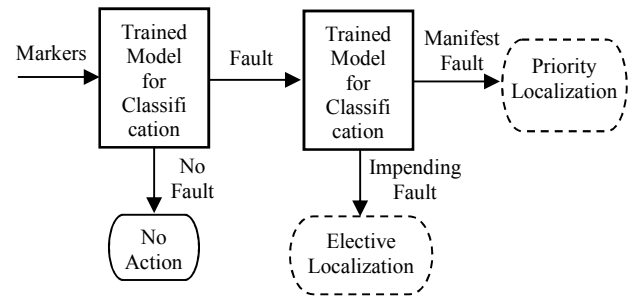

## F. Localization of Fault and Performance Conditions

The severity level of the faults would indicate whether they are warnings, minor, major or critical. In the case of major and critical faults, devices have degraded performance or have stopped working and need immediate action. Minor faults do not affect service and can be scheduled for localization accordingly. Warnings, along with the state information, provide insight into the degrading health of devices and could signal a major impending fault. We have evolved a multi-layer fault identification and localization system. At Layer-1, it detects the broad category of fault and then at Layer-2 does a fine-grain classification. In the case of impending faults, the system predicts the locations and severity levels of the developing faults. Table VI gives examples for each of the layers and associated markers.

TABLE VI. LIST OF PERFORMANCE AND FAULT ISSUES (NOT EXHAUSTIVE)

| Layer-1 | Layer-2 | Markers |
|---|---|---|
| Network Availability | • Base station distance<br>• Antenna height | Call Drop Rate (CDR), Signal Strength, Multi-Path Interference |
| | • Backhaul failure<br>• User equipment (UE) settings | Roaming not functional, GGSN alarm |
| | • Power module failure<br>• Radio unit failure | BTS downtime, No of affected cells Power failure alarm/notification |
| | • Insufficient TCH<br>• Locked Transceivers (TRXs) | Traffic Channel (TCH) congestion, blocking |
| Connection Maintenance | • Hopping Sequence No (HSN) clash<br>• LAC boundary | Handover success rate |
| Network Performance | • BCCH, TCH freq. plan<br>• Handover parameters<br>• Antenna tilt | Bad Rx quality |
| | • PGW fault<br>• Radio network fault | Packet Loss |
| Security | • DOS attack<br>• Home Subscriber Server (HSS) failure<br>• Record deletion | Tenant Authentication failure |
| Virtualization-Component Failure | • EM fault<br>• NE failure | Hardware failure alarm |
| Virtualization – Software vulnerabilities | • Physical resource congestion | Hypervisor alarm |

Both the machine learning models for detection use kernel based binary SVM. For localization of manifest faults initial tests have been carried out using SVM with multi-class modification and Random Forest. Localization of impending faults and performance issues has been implemented using stacked autoencoder – a deep neural network. Some of the major advantages of this algorithm are automatic selection of features from high-dimensional data and filtering information through the layers to achieve better accuracy. For training and testing our models, we adapted a composite feature dataset from the ones described in Section V(C). The implementation was done in MATLAB running on 2.7 GHz Intel Core i7 with 8 GB 1600 MHz DDR3 RAM and NVIDIA GeForce GT 650M 1024 MB graphics. In the next section, we present our evaluation results.

## VI. EVALUATION RESULTS

### A. Detection with shallow structures

Initial classification tests were carried out with SVM, ADT and Random Forests for binary 'fault', 'no-fault' classification. Each of the models was trained with 240 examples and 10% cross-validation. In the comparative test (Table VII) it can be seen that SVM performs better than the other methods with ≥ 95% accuracy. It can also be seen that true positive rate (TP) is high and false positive rate (FP) is low for the fault class 1 for SVM. High precision shows that no problem cases were correctly classified in SVM and RF.

TABLE VII COMPARATIVE RESULTS

| | SVM | ADT | Random Forests |
|---|---|---|---|
| Time taken | 0.01 sec | < 0.01 sec | 0.1 seconds |
| Correctly classified instances | 95.42% | 95.00% | 86.67% |
| Precision (Average) | 95.7% | 95.2% | 86.9% |
| Mean absolute error | 0.0458 | 0.0859 | 0.2509 |
| Root mean squared error | 0.2141 | 0.2092 | 0.3261 |
| True positive for class 0 | 94.3% | 94.3% | 95.5% |
| False positive for class 0 | 2.4% | 3.6% | 30.1% |
| True positive for class 1 | 97.6% | 96.4% | 69.9% |
| False positive for class 1 | 5.7% | 5.7% | 4.5% |

Selection of SVM was ratified by running it on the KDE dataset. The KDE dataset has been generated using kernel density estimation and Markov sampling. An extract can be seen in Table VIII. Only eight of the 26 features mentioned in Table 5 have been used to simplify data preparation and yet keep the data close to the real network. The data contains values of the selected features along with the severity of each fault case. The first model trained to detect faults (including warnings) given the feature values for all the new cases. It has to decide whether the given information indicates a possible fault or no detectable fault is there.

TABLE VIII EXTRACT OF TRAINING DATASET FOR DETECTION

| Docket # | C/I Ratio | Power Margin (dBm) | POI Cong (%) | CSSR (%) | TCH Cong (%) | SDCCH Cong (%) | Signal Strength dBm | Packet loss (%) | Severity |
|---|---|---|---|---|---|---|---|---|---|
| 23 | 49 | 22 | 6 | 99 | 0 | 0 | 96 | 0 | 2 |
| 52 | 10 | 13 | 8 | 83 | 10 | 0 | 109 | 2 | 1 |
| 68 | 28 | 20 | 1 | 96 | 1 | 1 | 100 | 1 | 0 |
| 69 | -10 | -15 | 11 | 91 | 6 | 0 | 98 | 0 | 1 |
| 134 | 67 | 25 | 14 | 85 | 3 | 4 | 80 | 2 | 1 |
| 201 | 49 | 15 | 1 | 98 | 2 | 0 | 95 | 2 | 0 |
| 215 | 49 | 75 | 6 | 99 | 5 | 0 | 75 | 0 | 3 |

The second model was trained to analyze the values of the features of all the faults and to decide whether there is a manifest fault or an impending fault. As discussed earlier, the features interact in a complex way to decide network performance. Not all warnings become faults in the future. The acid test for the predictive trained model was to examine whether it correctly predicts the risk of impending fault or performance issue that would need a truck roll out or other maintenance actions. The SVM model had nearly 100% accuracy of predicting the warning cases in our test sets.

We compared ours with fault detection in other areas. In [18] the authors used SVM to classify wind turbine faults using operational data and achieved 99.6% accuracy. In [27] wind turbine faults were detected with accuracy 98.26% for linear SVM and 97.35 for Gaussian. The authors in [19] studied faults in rotating machinery and got 99.9% accuracy of classification with SVM, higher than other methods studied.

*B. Localization with Deep Structure*

As mentioned before, we have implemented and tested localization of impending fault to predict where the impending fault might eventually happen and with what severity level. To evaluate the model we used the Telstra network fault datasets described in Section V(C). The training dataset has 7382 records, the feature dataset 58672 records and resource type 21077 records. Similarly, the other files are also fairly large. The feature set used for analysis is given in Table IX.

TABLE IX FEATURES FROM NETWORK FAULT DATASETS

| 1 | Id | 5 | Resource type 1 to 10 |
|---|---|---|---|
| 2 | Location | 6 | Severity type 1 to 5 |
| 3 | Features 1 to 386 | 7 | Event type |
| 4 | Volumes for features | 8 | Fault severity |

Fault_severity levels are- No fault (0), a few faults (1) and many faults (2) and are based on actual faults reported by users. Severity_type, on the other hand, describes the intensity of the warning, which can be used to predict impending faults. The stacked autoencoder was setup with 100 hidden layers in the first and 50 in the second autoencoder. The Softmax layer provides supervised back-propagation improvement of the weights learned during unsupervised training. A varying number of epochs (iterations) were used for different layers to obtain best results. Figure 5 gives a representation of the stacked autoencoder.

Fig 5 The stacked encoder used for prediction

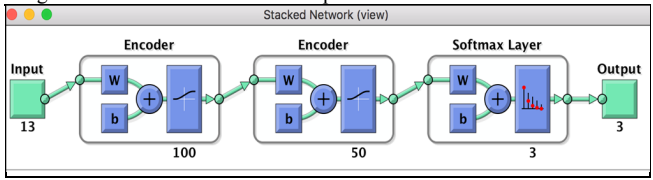

With default sparsity regulation of 1 the confusion matrix shows 97.3% accuracy of prediction (Fig 6) of severity levels of impending faults. With sparsity regularization parameter of 4 and sparsity proportion 0.1 (parameters arrived using grid search) the confusion matrix indicates 100% accuracy for the given training and test datasets (Fig. 7). The accuracy of the model is dependent on the size of the hidden layers compared to the input data. Also, the relative sizes of the hidden layers affect the accuracy. Fig. 8 shows the accuracy and mean square errors (MSE) for various sizes of the first hidden layer (H1) with the size of H2 as a parameter. The idea is to show that accuracy and MSE are good together for a certain range of H1 given the size of H2 and the parameters need to be carefully selected.

Comparing with results obtained by other researchers in studies of HVAC fault detection with deep learning deep belief network (DBN) 2 layer model the overall accuracy was claimed to be ≥95% [46]. Our results compare favorably with these results.

Fig. 6 Confusion matrix without sparsity regularization  Fig 7 Confusion matrix with sparsity regularization

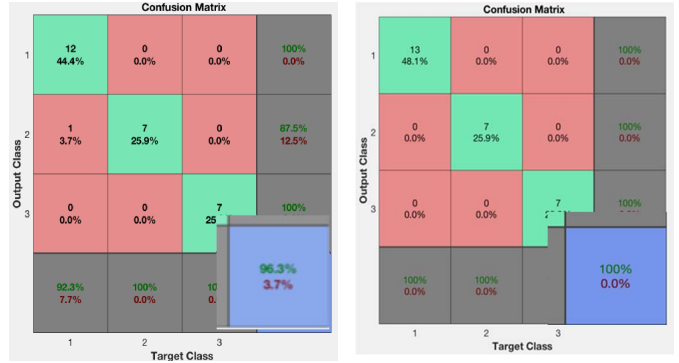

Fig. 7 Effect of hidden layers on accuracy and MSE

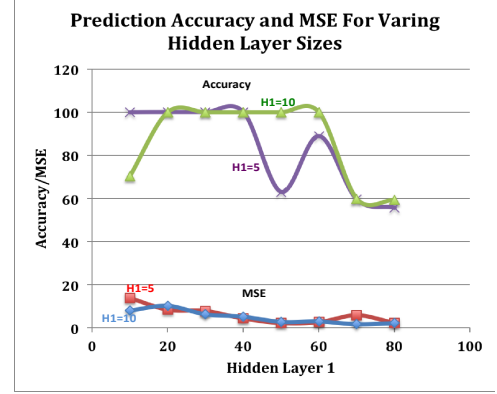

VII. SUMMARY

In this paper, a study of the markers and metrics related to fault and performance management of communication networks has been performed. Handling fault and performance anomalies when they occur is crucial for the success of NFV deployments over clouds. We propose a model for detection and localization of manifest and impending fault and performance issues. Some of the aspects of detection and localization of faults have been implemented using shallow and deep structures respectively. It has been observed that SVM for classification performs well for detection of fault/no-fault or manifest/impending fault situations. This information could then be used with localization function for deeper analysis of the warnings to predict impending faults and their severity. Deep structure of stacked autoencoder was used for careful examination of 'warning' cases to predict the severity of faults that may arise in future. This would help in planning resources for maintenance activities. The models have been evaluated with actual network and synthetic datasets with good results. It is proposed to carry out further work on fine grain localization for identification of resources, location and nature of the problem for more actionable information.